\documentclass[12pt,a4paper]{iopart}

\usepackage{iopams}
\usepackage{setstack}
\usepackage{graphicx}
\usepackage{pict2e}
\usepackage{color}
\usepackage{hyperref}

\newcommand{\sign}{\mathrm{sign}}
\newcommand{\Li}{\mathrm{Li}}

\newcommand{\dd}{\rmd}
\newcommand{\ee}{\rme}
\newcommand{\ii}{\rmi}

\renewcommand{\text}[1]{\mathrm{#1}}

\begin{document}
\title[Current fluctuations for the second class particle]{Current fluctuations for the second class particle : joint statistics}
\author{Sylvain Prolhac}
\address{Laboratoire de Physique Th\'eorique, Universit\'e de Toulouse, France}

\begin{abstract} We consider TASEP with a \emph{single} second class particle and periodic boundary conditions. Using Bethe ansatz, we compute stationary large deviations for the joint statistics of the current of first and second class particles. At large scales, the generating function of the joint cumulants shows an unexpected connection to current fluctuations of TASEP with open boundaries.\\\\
%	\keywords{TASEP, second class particle, Large deviations, Bethe ansatz}\\\\
\end{abstract}

%\red{\today}

\maketitle

\section{Introduction}
The one dimensional totally asymmetric simple exclusion process (TASEP) \cite{D1998.1,S2001.1,GM2006.1} is an exactly solvable model where particles randomly move with unit rate on a lattice, from one site to the next in the same direction, with hard-core repulsion preventing having multiple particles at the same site. At large scales, the model is known to belong to KPZ universality \cite{C2011.1,S2017.1,T2018.1}, in the sense that local currents of particles have the same statistics as the random field described by the KPZ fixed point, a universality class far from equilibrium with connections to interface growth \cite{KPZ1986.1,HHZ1995.1}, classical and quantum fluids \cite{vB2012.1,S2014.1,L2023.1,KL2013.1,PSMG2021.1} or random geometry \cite{G2022.1}.

Since KPZ fluctuations have long range correlations, boundary conditions matter when considering systems with a finite volume \cite{P2024.1}, and several results have been obtained so far for various kinds of statistics with periodic \cite{DL1998.1,P2016.1,BL2018.1} and open \cite{DEL2004.1,GLMV2012.1,P2024.2} boundary conditions. In this paper, we study the KPZ fixed point with periodic boundaries in the presence of a moving defect, which can be thought of as another kind of boundary condition, and is a first step toward studying coupled KPZ fields. At the level of TASEP, we study a model where, in addition to the usual (called first class hereafter) particles, a single second class particle is present, which moves on empty sites as a first class particle, but behaves as an empty site for first class particles.

On the hydrodynamic scale, second class particles are known to follow characteristics of the deterministic conservation equation for the density of first class particles \cite{F1992.1}, and have been much studied in connection with shocks and rarefaction fans \cite{DJLS1993.1,FF1994.1,FK1995.1,M1996.1,CP2013.1,FGN2019.1,KD2020.1,RV2022.1,CZ2024.1}. Current fluctuations of second class particles have also been investigated using Bethe ansatz for systems with periodic boundaries: in \cite{DE1999.1}, the diffusion constant of a single particle of second class was obtained, while in  \cite{C2008.1}, the average current of a system with an arbitrary number of second class particles was found. Both papers actually consider a more general case with arbitrary hopping rates $\alpha,\beta$ for the second class particles, while we restrict for simplicity to $\alpha=\beta=1$ here. Particles moving in both directions, with general asymmetry, have also been considered recently using Bethe ansatz \cite{LEM2024.1}.

We consider in this paper TASEP with $N$ first class particles and a single second class particle on a periodic lattice of $L$ sites (with $1\leq N\leq L-2$ to avoid degenerate cases). We are interested in the total time-integrated currents $Q_{1}(t)$ and $Q_{2}(t)$ of first and second class particles, incremented by $+1$ (respectively $-1$) each time a particle of the correct class moves forward (resp. backward) anywhere in the system, and initialized with $Q_{1}(0)=Q_{2}(0)=0$. Denoting first class particles by $1$, second class particles by $2$ and empty sites (holes) by $0$, the only transitions allowed on pairs $i,i+1$ of neighbouring sites, and the corresponding current increments, are
\begin{equation}
	\begin{array}{lll}
		10\to01 &\Rightarrow& Q_{1}\to Q_{1}+1\\
		20\to02 &\Rightarrow& Q_{2}\to Q_{2}+1\\
		12\to21 &\Rightarrow& Q_{1}\to Q_{1}+1\;\text{and}\;Q_{2}\to Q_{2}-1\;.
	\end{array}
\end{equation}
All these transitions happen with the same unit rate.

Since first class particles do not distinguish between the second class particle and a hole, and the second class particle moves on holes as a first class particle, $Q_{1}$ (respectively $Q_{1}+Q_{2}$) is simply equal to the current of $N$ (resp. $N+1$) TASEP particles in a single species system of $L$ sites with periodic boundaries. The main goal of the paper is then to compute the non-trivial correlations existing between $Q_{1}$ and $Q_{2}$, first for finite $L,N$, then under suitable large $L,N$ limit where KPZ fluctuations are expected.

Our starting point is the known \cite{DE1999.1,C2008.1} Bethe ansatz solution for the deformed generator $M(\gamma_{1},\gamma_{2})$ acting on the vector space generated by all possible configurations $C$ of the particles on the lattice, such that $\langle\ee^{\gamma_{1}Q_{1}(t)+\gamma_{2}Q_{2}(t)}\rangle=\sum_{C}\langle C|\ee^{tM(\gamma_{1},\gamma_{2})}|\mathbb{P}_{0}\rangle$ with $|\mathbb{P}_{0}\rangle$ the vector of initial probabilities. At late times $t\to\infty$, the generating function then behaves as
\begin{equation}
	\label{GF(Q1,Q2)}
	\langle\ee^{\gamma_{1}Q_{1}(t)+\gamma_{2}Q_{2}(t)}\rangle\asymp\ee^{t\,\lambda_{\rm st}(\gamma_{1},\gamma_{2})}\;,
\end{equation}
in the sense that the ratio of the logarithms converges to $1$ when $t\to\infty$. The eigenvalue $\lambda_{\rm st}(\gamma_{1},\gamma_{2})$ with largest real part of $M(\gamma_{1},\gamma_{2})$ then generates the joint stationary cumulants of $Q_{1}$ and $Q_{2}$. Our main results are the exact parametric expression (\ref{lambda[C,E]})-(\ref{gamma2[C,E]}) for finite $L,N$, obtained in section~\ref{section finite L}, and the large $L$ asymptotics (\ref{mu[chi]})-(\ref{s2[chi]}), obtained in section~\ref{section large L} (and extended analytically to all values of the parameters in \ref{appendix a.c.}). Technical details of the large $L$ asymptotics are presented in \ref{appendix large L}. An unexpected connection with expressions obtained in \cite{GLMV2012.1} for open TASEP is stated in (\ref{connection open TASEP}).

\section{Finite $L$ results by Bethe ansatz}
\label{section finite L}
The Bethe equations of TASEP with a single second class particle are \cite{DE1999.1,C2008.1} \footnote{Our notations are essentially the same as in \cite{DE1999.1}, with $M=N+1$ and $\gamma=\gamma_{2}$, for the special case with hopping rates $\alpha=\beta=1$ and $b=1-\beta=0$, and extended to non-zero $\gamma_{1}$. In the notations of \cite{C2008.1}, one has $N=L$, $M_{1}=N$, $M_{2}=1$, $\nu_{10}=\gamma_{1}$, $\nu_{20}=\gamma_{2}$, $\nu_{12}=\gamma_{1}-\gamma_{2}$, $Y_{j}=z_{j}$, $Z_{1}=E$, again for the special case $\alpha=\beta=1$, $b=1-\beta=0$. Note that there is a misprint in the published version of \cite{C2008.1}, with a factor $(-1)^{M_{1}}$ missing in the Bethe equation (36) there.}
\begin{equation}
	\label{Bethe eq}
	C\,z_{j}^{L}\,\Big(\frac{1}{1-z_{j}}-E\Big)+(z_{j}-1)^{N}=0\;,
\end{equation}
$j=0,\ldots,N$, where $C$ and $E$ are related to the conjugate variables $\gamma_{1}$ and $\gamma_{2}$ through
\begin{eqnarray}
	\label{gamma1[zj]}
	&&\ee^{L\gamma_{1}}+C\prod_{j=0}^{N}\frac{1}{1-z_{j}}=0\\
	\label{gamma2[zj]}
	&&\ee^{L\gamma_{2}}+C\prod_{j=0}^{N}\Big(\frac{1}{1-z_{j}}-E\Big)=0\;.
\end{eqnarray}
Given a solution of the Bethe equation, the corresponding eigenvalue $\lambda$ of $M(\gamma_{1},\gamma_{2})$ is then equal to
\begin{equation}
	\label{lambda[zj]}
	\lambda=\sum_{j=0}^{N}\Big(\frac{1}{z_{j}}-1\Big)\;.
\end{equation}

By the Perron-Frobenius theorem, the eigenvalue $\lambda_{\rm st}(\gamma_{1},\gamma_{2})$ of $M(\gamma_{1},\gamma_{2})$ with largest real part is non-degenerate and analytic for all real values of $\gamma_{1},\gamma_{2}$, and coincides with the stationary eigenvalue $\lambda_{\rm st}(0,0)=0$ when $\gamma_{1}=\gamma_{2}=0$. The corresponding solution of the Bethe equation is selected by the requirement
\begin{equation}
	\label{selection st}
	z_{j}\to1
\end{equation}
for all $j=0,\ldots,N$ when $\gamma_{1},\gamma_{2}\to0$, corresponding in the variables $C,E$ to $C\to0$ with $E$ depending on the ratio $\gamma_{1}/\gamma_{2}$.

The sums and products over $j$ in (\ref{gamma1[zj]})-(\ref{lambda[zj]}) can be computed from residues, using
\begin{equation}
	\sum_{j=0}^{N}h(z_{j})=\oint_{\mathcal{C}}\dd z\,h(z)\,\frac{R'(z)}{R(z)}
\end{equation}
with $R(z)=C\,z^{L}\,(1-E(1-z))-(z-1)^{N+1}$ the polynomial appearing in the Bethe equations (\ref{Bethe eq}), and $\mathcal{C}$ a simple closed curve with positive orientation such that the $N+1$ Bethe roots $z_{j}$ for the stationary eigenstate are surrounded by $\mathcal{C}$ while the $L-N$ other roots of $R$ and the poles of $h$ are outside $\mathcal{C}$. Following \cite{DL1998.1,DE1999.1}, a series expansion in the variable $C$ allows to compute the contour integral for the Bethe eigenstate satisfying (\ref{selection st}) by evaluating the residue at $z=1$. Rather straightforward calculations then lead to the following parametric representation
\begin{eqnarray}
	\label{lambda[C,E]}
	&&\lambda_{\rm st}=-\sum_{m=1}^{\infty}\frac{C^{m}}{m}\sum_{p=0}^{m}E^{p}{{m}\choose{p}}{{Lm-2}\choose{Nm+m-p-1}}\\
	\label{gamma1[C,E]}
	&&L\gamma_{1}=-\sum_{m=1}^{\infty}\frac{C^{m}}{m}\sum_{p=0}^{m}E^{p}{{m}\choose{p}}{{Lm}\choose{Nm+m-p}}\\
	\label{gamma2[C,E]}
	&&L\gamma_{2}=-\sum_{m=1}^{\infty}\frac{C^{m}}{m}\sum_{p=0}^{m}E^{p}{{m-1}\choose{p}}{{Lm}\choose{Nm+m-p}}\;.
\end{eqnarray}

The first joint cumulants of the currents $Q_{1}$ and $Q_{2}$, which are the derivatives of $\lambda_{\rm st}(\gamma_{1},\gamma_{2})$ at $\gamma_{1}=\gamma_{2}=0$, can be computed by inverting (\ref{gamma1[C,E]}) and (\ref{gamma2[C,E]}) perturbatively in $C$ and $E$ and injecting into (\ref{lambda[C,E]}). Computing finally the large $L$ asymptotics at finite density $\rho\in(0,1)$ with
\begin{equation}
	\rho=N/L\;,
\end{equation}
one recovers the known stationary averages
\begin{eqnarray}
	\label{J1st}
	&&J_{1}^{\rm st}=\lim_{t\to\infty}\frac{\langle Q_{1}(t)\rangle}{t}=\frac{N(L-N)}{L-1}\simeq\rho(1-\rho)L\\
	\label{J2st}
	&&J_{2}^{\rm st}=\lim_{t\to\infty}\frac{\langle Q_{2}(t)\rangle}{t}=\frac{L-2N-1}{L-1}\simeq1-2\rho
\end{eqnarray}
and variances \cite{DEM1993.1,DE1999.1}
\begin{eqnarray}
	&&\fl\lim_{t\to\infty}\frac{\mathrm{Var}(Q_{1}(t))}{t}=\frac{LN(L-N)}{(L-1)(2L-1)}\,\frac{{{2L}\choose{2N}}}{{{L}\choose{N}}^{2}}\simeq\frac{\sqrt{\pi}\,(\rho(1-\rho)L)^{3/2}}{2}\\
	&&\fl\lim_{t\to\infty}\frac{\mathrm{Var}(Q_{2}(t))}{t}=\frac{L(L-5)N(L-N-1)+L(L-1)(2L-1)}{(L-1)(2L-1)(2N+1)(L-N)}\,\frac{{{2L}\choose{2N}}}{{{L}\choose{N}}^{2}}\simeq\frac{\sqrt{\pi\rho(1-\rho)L}}{4}\;.\nonumber
\end{eqnarray}
Additionally, the stationary covariance of $Q_{1}$ and $Q_{2}$ is found to be equal to
\begin{eqnarray}
	\lim_{t\to\infty}\frac{\langle Q_{1}Q_{2}\rangle-\langle Q_{1}\rangle\langle Q_{2}\rangle}{t}&&=\frac{LN(L-3N-2)}{(L-1)(2L-1)(2N+1)}\,\frac{{{2L}\choose{2N}}}{{{L}\choose{N}}^{2}}\nonumber\\
	&&\simeq\frac{(1-3\rho)\sqrt{\pi\rho(1-\rho)L}}{4}
\end{eqnarray}

\section{Large $L$ asymptotics}
\label{section large L}
A proper large $L$ asymptotics of the exact parametric expression (\ref{lambda[C,E]})-(\ref{gamma2[C,E]}) for the joint cumulant generating function requires scaling the variables $C$ and $E$ as
\begin{eqnarray}
	\label{C[v]}
	&&C=-\sqrt{\rho(1-\rho)L}\,(\rho^{\rho}(1-\rho)^{1-\rho})^{L}\,\frac{\rho\,v}{1-\rho}\\
	\label{E[u/v]}
	&&E=-\frac{1-\rho}{\rho}\,\exp\Big(\frac{u/v}{\sqrt{\rho(1-\rho)L}}\Big)\;,
\end{eqnarray}
with $\rho=N/L$ the density of first class particles, and new variables $u$ and $v$ independent of $L$. Then, introducing the scaled variables
\begin{eqnarray}
	\label{t[tau]}
	t=\frac{\tau L^{3/2}}{\sqrt{\rho(1-\rho)}}\\
	\label{gamma1[s1]}
	\gamma_{1}=\frac{s_{1}}{\sqrt{\rho(1-\rho)}\,L^{3/2}}\\
	\label{gamma2[s2]}
	\gamma_{2}=\frac{s_{2}}{L}\\
	\label{lambda[mu]}
	\lambda_{\rm st}(\gamma_{1},\gamma_{2})=J_{1}^{\rm st}\,\gamma_{1}+J_{2}^{\rm st}\,\gamma_{2}+\frac{\sqrt{\rho(1-\rho)}}{L^{3/2}}\,\mu_{\rm st}(s_{1},s_{2})\;,
\end{eqnarray}
with $J_{1}^{\rm st}$, $J_{2}^{\rm st}$ the stationary averages (\ref{J1st}), (\ref{J2st}), and defining the current fluctuations
\begin{equation}
	\label{xi[Q]}
	\xi_{1}(\tau)=\frac{Q_{1}(t)-t\,J_{1}^{\rm st}}{\sqrt{\rho(1-\rho)}\,L^{3/2}}
	\quad\text{and}\quad
	\xi_{2}(\tau)=\frac{Q_{2}(t)-t\,J_{2}^{\rm st}}{L}\;,
\end{equation}
a computation sketched in \ref{appendix large L} shows that the generating function (\ref{GF(Q1,Q2)}) takes at large $L$ and $\tau\gg1$ the form
\begin{equation}
	\label{GF(xi1,xi2)}
	\langle\ee^{s_{1}\,\xi_{1}(\tau)+s_{2}\,\xi_{2}(\tau)}\rangle\asymp\ee^{\tau\,\mu_{\rm st}(s_{1},s_{2})}\;.
\end{equation}
Introducing the function
\begin{equation}
	\label{chi sum}
	\chi(u,v)=-\frac{1}{\sqrt{2\pi}}\sum_{{j,k=0}\atop{(j,k)\neq(0,0)}}^{\infty}\frac{(j+2k)!}{(j+2k)^{k+5/2}}\,\frac{u^{j}}{j!}\,\frac{(-v^{2}/2)^{k}}{k!}\;,
\end{equation}
defined in some neighbourhood of $(u,v)=(0,0)$, one finds in terms of the partial derivatives $\partial_{u}=\frac{\partial}{\partial u}$ and $\partial_{v}=\frac{\partial}{\partial v}$ of the function $\chi(u,v)$ the parametric expression
\begin{eqnarray}
	\label{mu[chi]}
	\mu_{\rm st}(s_{1},s&_{2})&=(1-u\partial_{u})\chi(u,v)\\
	\label{s1[chi]}
	&s_{1}&=(u\partial_{u}+v\partial_{v})\chi(u,v)\\
	\label{s2[chi]}
	&s_{2}&=-v\partial_{u}\chi(u,v)\;,
\end{eqnarray}
valid in a neighbourhood of $(s_{1},s_{2})=(0,0)$. Analytic continuation to all $(s_{1},s_{2})\in\mathbb{R}^{2}$ is performed explicitly in \ref{appendix a.c.}.

We emphasize that the final result for $\mu_{\rm st}(s_{1},s_{2})$ is independent of the density of first class particles $\rho$. This is a manifestation of the universality of KPZ fluctuations, where limiting random processes $\xi_{1}(\tau)$ and $\xi_{2}(\tau)$ obtained after an appropriate $\rho$-dependent centering and scaling become independent of that parameter after taking the limit $L\to\infty$. Additionally, the independence of $\xi_{1}(\tau)$ and $\xi_{2}(\tau)$ from $\rho$ is expected to hold at any finite time $\tau$ as well, and not just at the level of stationary large deviations.

\begin{table}
	\begin{center}
		\begin{tabular}{c|ccccc}
			\raisebox{-1mm}{$p$}$\,\backslash\,$\raisebox{1mm}{$q$} & $0$ & $1$ & $2$ & $3$ & $4$\\[1mm]\hline
			\\[-3mm]
			$0$ & $0$ & $0$ & $1/4$ & $0$ & $-\frac{3}{8}-\frac{9}{16\sqrt{2}}+\frac{4}{3\sqrt{3}}$ \\[5mm]
			$1$ & $0$ & $0$ & $0$ & $0$ & $0$ \\[5mm]
			$2$ & $1/2$ & $0$ & $-\frac{5}{8}-\frac{3}{4\sqrt{2}}+\frac{2}{\sqrt{3}}$ & $0$ & \begin{tabular}{r}$\frac{755}{36}+\frac{387}{16\sqrt{2}}-\frac{104}{3\sqrt{3}}$\\$-\frac{576}{25\sqrt{5}}-\frac{19}{\sqrt{6}}$\end{tabular} \\[7mm]
			$3$ & $\frac{3}{2}-\frac{8}{3\sqrt{3}}$ & $0$ & $-\frac{317}{36}-\frac{12}{\sqrt{2}}+\frac{24}{\sqrt{3}}+\frac{192}{25\sqrt{5}}$ & $0$ & \begin{tabular}{r}$\frac{8095}{16}+\frac{1701}{4\sqrt{2}}-\frac{15278}{27\sqrt{3}}-\frac{11664}{25\sqrt{5}}$\\$-\frac{772}{\sqrt{6}}-\frac{34560}{343\sqrt{7}}+\frac{7936}{25\sqrt{15}}$\end{tabular} \\[7mm]
			$4$ & $\frac{15}{2}+\frac{9}{\sqrt{2}}-\frac{24}{\sqrt{3}}$ & $0$ & \begin{tabular}{r}$-\frac{1687}{12}-\frac{315}{2\sqrt{2}}+\frac{240}{\sqrt{3}}$\\$+\frac{144}{\sqrt{5}}+\frac{120}{\sqrt{6}}$\end{tabular} & $0$ &
			\begin{tabular}{l}$\frac{489905}{48}+\frac{818135}{96\sqrt{2}}-\frac{93520}{9\sqrt{3}}-\frac{7992}{\sqrt{5}}$\\$-\frac{19875}{\sqrt{6}}-\frac{155520}{49\sqrt{7}}-\frac{11016}{5\sqrt{10}}+\frac{64896}{5\sqrt{15}}$\end{tabular}
		\end{tabular}
	\end{center}
	\caption{Exact values of first $c_{p,q}/\pi^{\frac{p+q-1}{2}}$, with joint cumulants $c_{p,q}$ defined in (\ref{cpq}).}
	\label{table joint cumulants exact}
\end{table}

\begin{table}
	\begin{center}
		\begin{tabular}{c|ccccc}
			\raisebox{-1mm}{$p$}$\,\backslash\,$\raisebox{1mm}{$q$} & $0$ & $1$ & $2$ & $3$ & $4$\\[1mm]\hline
			\\[-4mm]
			$0$ & $0$ & $0$ & $0.443113$ & $0$ & $-0.0164110$ \\
			$1$ & $0$ & $0$ & $0$ & $0$ & $0$ \\
			$2$ & $0.886227$ & $0$ & $-0.00350553$ & $0$ & $0.000689260$ \\
			$3$ & $-0.124409$ & $0$ & $0.00167728$ & $0$ & $-0.000343080$ \\
			$4$ & $0.0420663$ & $0$ & $-0.000597524$ & $0$ & $0.0000716995$
		\end{tabular}
	\end{center}
	\caption{Numerical values of the first joint cumulants $c_{p,q}$ defined in (\ref{cpq}).}
	\label{table joint cumulants num}
\end{table}

By definition (\ref{xi[Q]}), the late time averages of $\xi_{1}$ and $\xi_{2}$ vanish,
\begin{equation}
	\lim_{\tau\to\infty}\frac{\langle\xi_{1}(\tau)\rangle}{\tau}=0
	\quad\text{and}\quad
	\lim_{\tau\to\infty}\frac{\langle\xi_{2}(\tau)\rangle}{\tau}=0\;,
\end{equation}
which is recovered from the exact solution by solving perturbatively $(u,v)$ as a function of $(s_{1},s_{2})$ using (\ref{s1[chi]}), (\ref{s2[chi]}) and injecting into (\ref{mu[chi]}). Pushing the expansions further, we recover the known variances stated in the previous section,
\begin{equation}
	\lim_{\tau\to\infty}\frac{\mathrm{Var}(\xi_{1}(\tau))}{\tau}=\frac{\sqrt{\pi}}{2}
	\quad\text{and}\quad
	\lim_{\tau\to\infty}\frac{\mathrm{Var}(\xi_{2}(\tau))}{\tau}=\frac{\sqrt{\pi}}{4}\;.
\end{equation}
Higher cumulants of $\xi_{1}(\tau)$ are obtained by setting $s_{2}=0$, which corresponds to $v=0$, and we recover as expected the polylogarithm $\chi(u,0)=-\frac{1}{\sqrt{2\pi}}\,\Li_{5/2}(u)$ (with $\Li_{5/2}(u)=\sum_{j=1}^{\infty}\frac{u^{j}}{j^{5/2}}$) appearing for periodic TASEP without second class particles \cite{DL1998.1}. Higher cumulants of $\xi_{2}(\tau)$ are new. They are obtained at $s_{1}=0$, which corresponds to a non-trivial curve in $(u,v)$-plane with $u\simeq\frac{v^{2}}{4\sqrt{2}}$ when $(u,v)\to(0,0)$.

\begin{figure}
	\begin{center}
		\includegraphics[width=75mm]{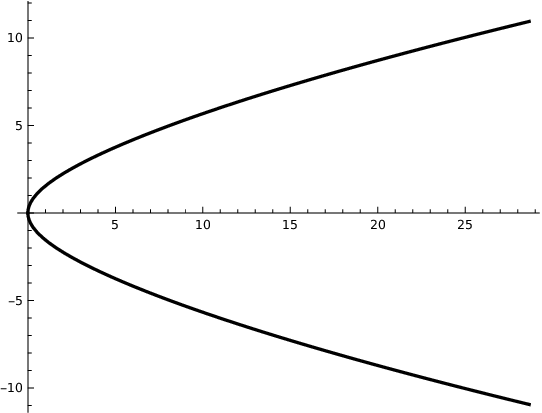}
		\begin{picture}(0,0)
			\put(-5,29.5){$s_{1}$}
			\put(-71,55){$s_{2}$}
		\end{picture}
	\end{center}
	\caption{Curve in $(s_{1},s_{2})$-plane where a connection to open TASEP is observed, plotted from the parametric expression $(s_{1}(v),s_{2}(v))$ with $s_{1}(v)$ and $s_{2}(v)$ given below (\ref{connection open TASEP}), evaluated numerically using the analytically continued integral expression (\ref{chi int log}) for $\chi(u,v)$.}
	\label{fig s1 s2 open}
\end{figure}
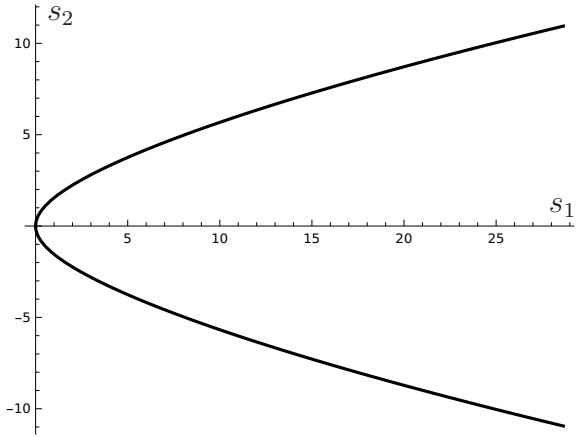

Additionally, on the line $u=0$ in $(u,v)$-plane, we observe an unexpected connection to stationary large deviations of \emph{open TASEP} (without second class particles), in the maximal current phase corresponding in KPZ language to infinite boundary slopes $\partial_{x}h$ for the random height field $h(x,\tau)$, see e.g. \cite{P2024.1}. Indeed, considering the time integrated current $Y_{t}^{\rm open}$ of particles entering the system up to time $t$ for open TASEP on $L$ sites with particles entering the system on the first site with rate $\alpha>1/2$ and leaving the system from site $L$ with rate $\beta>1/2$, and its fluctuations $\xi^{\rm open}(\tau)$ on the KPZ time scale $t=2\tau L^{3/2}$ defined from $Y_{t}^{\rm open}=\frac{t}{4}+\frac{\xi^{\rm open}(\tau)}{2}\sqrt{L}$, comparison with the results from \cite{GLMV2012.1} gives
\begin{equation}
	\label{connection open TASEP}
	\lim_{\tau\to\infty}\frac{\log\langle\ee^{s_{1}(v)\,\xi^{\rm open}(\tau)}\rangle}{\tau}
	=\chi(0,v)
	=\lim_{\tau\to\infty}\frac{\log\langle\ee^{s_{1}(v)\,\xi_{1}(\tau)+s_{2}(v)\,\xi_{2}(\tau)}\rangle}{\tau}\;,
\end{equation}
where $s_{1}(v)=v\partial_{v}\chi(0,v)$ and $s_{2}(v)=-v\partial_{u}\chi(0,v)$ with $\chi$ defined in (\ref{chi sum}). On the corresponding curve in $(s_{1},s_{2})$-plane, plotted in figure~\ref{fig s1 s2 open}, one has the asymptotics $s_{1}\simeq\frac{8}{3\pi}\,|s_{2}|^{3/2}$ and $\mu_{\rm st}(s_{1},s_{2})=\chi(0,v)\simeq\frac{(3\pi)^{2/3}}{5}\,s_{1}^{5/3}$ when $s_{2}\to\pm\infty$ (obtained from the integral expression (\ref{chi int log}) for $\chi(u,v)$, valid when $u<1$ and $v\neq0$). Expanding (\ref{connection open TASEP}) in powers of $v$ leads to identities between stationary averages, such as
\begin{equation}
	\lim_{\tau\to\infty}\frac{\langle\xi^{\rm open}(\tau)\rangle-\langle\xi_{1}(\tau)\rangle}{\tau}
	=\frac{2}{\sqrt{\pi}}\,\lim_{\tau\to\infty}\frac{\mathrm{Var}(\xi_{2}(\tau))}{\tau}\;.
\end{equation}
The observation (\ref{connection open TASEP}) suggests a possible relationship for KPZ fluctuations between the kind of moving boundary induced by the second class particle in a periodic system and true open boundary conditions with fixed density of particles at the edge. This observation is also reminiscent \footnote{We thank K. Mallick for pointing this out to us.} of the similarity between the matrix product representation for the stationary state of open TASEP and periodic TASEP with a single second class particle \cite{DEHP1993.1}, which corresponds at the level of KPZ fluctuations to roughly similar (yet distinct) Brownian measures for the stationary height field.

Joint stationary cumulants
\begin{equation}
	\label{cpq}
	c_{p,q}=\mu_{\rm st}^{(p,q)}(s_{1}=0,s_{2}=0)
	=\lim_{\tau\to\infty}\frac{\partial_{s_{1}}^{p}\partial_{s_{2}}^{q}\log(\langle\ee^{s_{1}\,\xi_{1}(\tau)+s_{2}\,\xi_{2}(\tau)}\rangle)_{|s_{1}=s_{2}=0}}{\tau}
\end{equation}
of $\xi_{1}(\tau)$ and $\xi_{2}(\tau)$ can be extracted from (\ref{mu[chi]})-(\ref{s1[chi]}) as well, see tables~\ref{table joint cumulants exact} and \ref{table joint cumulants num} for exact and numerical values of the first $c_{p,q}$. This confirms in particular that the random variables $\xi_{1}(\tau)$ and $\xi_{2}(\tau)$ are not independent even at late times: while
\begin{equation}
	c_{1,1}=\lim_{\tau\to\infty}\frac{\langle\xi_{1}(\tau)\,\xi_{2}(\tau)\rangle-\langle\xi_{1}(\tau)\rangle\,\langle\xi_{2}(\tau)\rangle}{\tau}=0\;,
\end{equation}
and more generally $c_{p,q}=0$ when $p=1$ or $q$ is odd, one finds
\begin{equation}
	\fl\hspace*{15mm}
	c_{2,2}=\lim_{\tau\to\infty}\frac{\langle\xi_{1}^{2}\,\xi_{2}^{2}\rangle-\langle\xi_{1}^{2}\rangle\,\langle\xi_{2}^{2}\rangle}{\tau}=\Big(\frac{2}{\sqrt{3}}-\frac{3}{4\sqrt{2}}-\frac{5}{8}\Big)\,\pi^{3/2}\approx-0.00350553\;,
\end{equation}
where several terms in the expression of $c_{2,2}$ in terms of moments of $\xi_{1}$ and $\xi_{2}$ were removed using $\partial_{s_{1}}^{p}\partial_{s_{2}}^{q}\log(\langle\ee^{s_{1}\,\xi_{1}(\tau)+s_{2}\,\xi_{2}(\tau)}\rangle)_{|s_{1}=s_{2}=0}=\mathcal{O}(\tau^{0})$ when $c_{p,q}=0$. Remarkably, we observe that the cancellation between the positive and negative terms in the exact expression for $c_{2,2}$ is almost perfect. This is also the case for higher cumulants, compare tables~\ref{table joint cumulants exact} and \ref{table joint cumulants num}.

\section{Conclusions}
We obtained stationary large deviations (\ref{lambda[C,E]})-(\ref{gamma2[C,E]}) for the joint statistics of first and second class particles of TASEP with periodic boundaries, when a single second class particle is present in the system. In the KPZ scaling limit, the joint cumulant generating function takes a special parametric form (\ref{mu[chi]})-(\ref{s2[chi]}) involving a function $\chi(u,v)$ and its partial derivatives.

As noted earlier for similar expressions obtained for TASEP with only first class particles and periodic or open boundaries, this special parametric form is reminiscent of the result of a saddle point approximation for an integral, and suggests the existence of a missing overarching principle for stationary large deviations of KPZ fluctuations. In particular, a physical interpretation for the variables $u$ and $v$, which appear for the moment as mere by-products of the Bethe ansatz solution, would be welcome.

Additionally, we observed here an intriguing coincidence (\ref{connection open TASEP}) between the function $\chi(0,v)$ and the analogue of $\chi$ for open TASEP with only first class particles and large enough boundary rates. It would be interesting to see whether this coincidence persists for the case of a second class particle with general hopping rates $\alpha,\beta$, as considered in \cite{DE1999.1,C2008.1}, when comparing to open TASEP with general boundary rates.

Another natural extension of our results would be late time corrections to stationary large deviations, which feature the last observable dependency on the initial state, and were computed for periodic TASEP \cite{MP2018.1} and open TASEP \cite{P2024.2}. The same kind of methods used there would seem to naturally extend in the presence of a second class particle.

\appendix
\section{Large $L$ asymptotics}
\label{appendix large L}
A naive look at the large $L$ behaviour of the binomial coefficients in (\ref{lambda[C,E]})-(\ref{gamma2[C,E]}) would suggest scaling $C$ simply as $(\rho^{\rho}(1-\rho)^{1-\rho})^{L}$ at large $L$ with fixed density $\rho=N/L$. It turns out, however, that the sum over $p$ would then cancel for each $m$ all the divergent terms of the large $L$ expansion because of the identity
\begin{equation}
	\label{sum bin 0}
	\sum_{p=0}^{m}{{m}\choose{p}}(-1)^{p}\,p^{r}=0
	\quad\text{for\;all}\;r=0,\ldots,m-1\;,
\end{equation}
which follows from evaluating the $r$-th derivative with respect to $z$ of the expansion of $(1-z)^{m}$ by the binomial theorem. An extra factor $\sqrt{L}$ is then needed for $C$, as specified in (\ref{C[v]}).

The large $L$ expansion of the binomial coefficients in (\ref{lambda[C,E]})-(\ref{gamma2[C,E]}) then requires in principle the full Stirling expansion of the factorial in terms of Bernoulli numbers,
\begin{equation}
	\label{Stirling}
	M!\simeq(M/\ee)^{M}\sqrt{2\pi M}\,\exp\Bigg(\sum_{k=1}^{\infty}\frac{(-1)^{k+1}B_{k+1}}{k(k+1)M^{k}}\Bigg)\;.
\end{equation}
However, because of the identity (\ref{sum bin 0}), most contributions to the binomial coefficient cancel after summation over $p$ using (\ref{sum bin 0}), and it turns out that the Bernoulli corrections in (\ref{Stirling}) do not contribute after summation over $p$. In the end, the large $L$ asymptotics with constant $\rho=N/L$, $m$, $p$, $a$, $b$ (such that $0<\rho<1$ and $m>0$)
\begin{eqnarray}
	\label{binomial asymptotics}
	&&\fl{{Lm-a}\choose{Nm+m-p-b}}\simeq\frac{\rho^{p-m+b}\,(1-\rho)^{m-p+a-b}}{\sqrt{2\pi\rho(1-\rho)mL}\,(\rho^{\rho}(1-\rho)^{1-\rho})^{mL}}\\
	&&\fl\hspace*{13mm}\times\Big[\ee^{-\frac{(m-p)^{2}}{2\rho(1-\rho)mL}}\Big(1-\frac{(1-2\rho(1-a)-2b)\,(m-p)}{2\rho(1-\rho)mL}+\frac{(1-2\rho)\,(m-p)^3}{6\rho^{2}(1-\rho)^{2}m^{2}L^{2}}\Big)+\ldots\Big]\;,\nonumber
\end{eqnarray}
where $\ldots$ is a term of the form $\sum_{k=0}^{\infty}Q_{2k-2}(m-p)/L^{k}$ with $Q_{2k-2}$ a polynomial of degree at most $2k-2$ (whose coefficients may depend on $\rho$, $a$ and $b$), is enough to extract the large $L$ asymptotics (\ref{chi sum})-(\ref{s2[chi]}) of (\ref{gamma1[s1]})-(\ref{lambda[mu]}) under the scalings (\ref{C[v]}), (\ref{E[u/v]}) for $C$ and $E$. In the case of (\ref{lambda[mu]}), combining the binomial coefficients coming from $\lambda_{\rm st}$, $\gamma_{1}$ and $\gamma_{2}$ is required before using (\ref{binomial asymptotics}).

\section{Analytic continuation of $\chi(u,v)$}
\label{appendix a.c.}
The parametric representation (\ref{chi sum})-(\ref{s2[chi]}) for $\mu(s_{1},s_{2})$, written in terms of variables $u$ and $v$, is only valid in a neighbourhood of $(u,v)=(0,0)$, corresponding to $(s_{1},s_{2})=(0,0)$. In this section, assuming (real) analyticity in $s_{1}$ and $s_{2}$, we extend $\mu(s_{1},s_{2})$ to all $(s_{1},s_{2})\in\mathbb{R}^{2}$. The desired analytic extension, which is not single valued in the variables $u$ and $v$, requires an extended domain built from pieces of the $(u,v)$-plane glued together.

\subsection{Integral formula for $\chi(u,v)$}
In a neighbourhood of $(u,v)=(0,0)$, the function $\chi(u,v)$ defined in (\ref{chi sum}) coincides with
\begin{equation}
	\label{chi int log}
	\chi(u,v)=-\int_{-\infty}^{\infty}\frac{\dd w}{2\pi}\,\Big(1-\frac{uw}{\ii v}-w^{2}\Big)\log\Big(1-\frac{u+\ii vw}{\ee^{w^{2}/2}}\Big)\;.
\end{equation}
This follows directly by expanding the logarithm in (\ref{chi int log}) and performing the Gaussian integrations on $w$. For convenience, we eliminate the logarithm in (\ref{chi int log}) by partial integration, leading in a neighbourhood of $(u,v)=(0,0)$ to $\chi(u,v)=\hat{\chi}(u,v)$ with
\begin{equation}
	\label{chi hat (u,v)}
	\hat{\chi}(u,v)=\int_{-\infty}^{\infty}\frac{\dd w}{2\pi}\,\frac{(w-\frac{uw^{2}}{2\ii v}-\frac{w^{3}}{3})(uw+\ii v(w^{2}-1))}{\ee^{w^{2}/2}-u-\ii vw}\;.
\end{equation}
The integral in (\ref{chi hat (u,v)}) is well defined except if the denominator $\ee^{w^{2}/2}-u-\ii vw$ vanishes for some $w\in\mathbb{R}$, which happens either when $u=1$ (at $w=0$) or when $v=0$ and $u\geq1$ (at $w=\pm\sqrt{2\log u}$). The expression (\ref{chi hat (u,v)}) for $\chi(u,v)$ is thus valid whenever $u\leq1$, and $v\neq0$ because of the term $\frac{uw^{2}}{2\ii v}$ in the numerator. When $v\to0$, the integrand has a divergent term $\propto v^{-1}$ which is odd with respect to $w$ and whose integral over $v$ vanishes, leaving the well defined limit when $u\leq1$
\begin{equation}
	\label{chi hat (u)}
	\hat{\chi}(u)=\lim_{v\to0}\hat{\chi}(u,v)=-\int_{-\infty}^{\infty}\frac{\dd w}{6\pi}\,\frac{uw^{4}}{\ee^{w^{2}/2}-u}
\end{equation}
obtained after subtracting a term proportional to $\frac{\dd}{\dd w}\frac{w^{3}}{\ee^{w^{2}/2}-u}$ in the integrand, which is a standard integral representation for the polylogarithm $\hat{\chi}(u)=-\frac{1}{\sqrt{2\pi}}\Li_{5/2}(u)$.

\begin{figure}
	\begin{center}
		\begin{tabular}{ll}
			\begin{picture}(75,105)
				\put(77,-5){\color[rgb]{0.7,0.7,0.7}\line(0,1){110}}
				\put(29,102){Case $v>0$}
				\put(0,90){
					\put(0,0){\thicklines\line(1,0){75}}
					\put(60,0){\thicklines\vector(1,0){1}}
					\put(72,1){$w$}
					\put(0,-10){$u<1$}
					\put(40,-10){\color{blue}\vector(0,1){7}}
					\put(40,-10){\circle*{2}}
					\put(42,-11){$w_{*}(u,v)$}
					\put(39,-25){\color{blue}\rotatebox[origin=c]{-90}{$\longrightarrow$}}
					\put(0,-23){analytic continuation}
					\put(8,-27){when $u$ crosses $1$}
				}
				\put(0,35){
					\put(0,0){\thicklines\line(1,0){10}}
					\put(30,0){\thicklines\qbezier(-20,0)(-12.5,0)(-5,7.5)}
					\put(30,0){\thicklines\qbezier(-5,7.5)(2.5,15)(10,15)}
					\put(40.5,15){\thicklines\vector(1,0){1}}
					\put(30,0){\thicklines\qbezier(10,15)(17.5,15)(25,7.5)}
					\put(30,0){\thicklines\qbezier(25,7.5)(32.5,0)(40,0)}
					\put(70,0){\thicklines\line(1,0){5}}
					\put(72,1){$w$}
					\put(0,10){$u>1$}
					\put(40,10){\circle*{2}}
					\put(33,5){$w_{*}(u,v)$}
					\put(39.5,-10){\rotatebox[origin=c]{90}{$=$}}
				}
				\put(0,0){
					\put(0,0){\thicklines\line(1,0){75}}
					\put(40,10){\thicklines\circle{10}}
					\put(40.5,15){\thicklines\vector(1,0){1}}
					\put(60,0){\thicklines\vector(1,0){1}}
					\put(72,1){$w$}
					\put(40,10){\circle*{2}}
					\put(46,9){$w_{*}(u,v)$}
				}
			\end{picture}
			&
			\begin{picture}(75,105)
				\put(29,102){Case $v<0$}
				\put(0,80){
					\put(0,0){\thicklines\line(1,0){75}}
					\put(60,0){\thicklines\vector(1,0){1}}
					\put(72,1){$w$}
					\put(0,10){$u<1$}
					\put(40,10){\color{blue}\vector(0,-1){7}}
					\put(40,10){\circle*{2}}
					\put(42,9){$w_{*}(u,v)$}
					\put(39,-15){\color{blue}\rotatebox[origin=c]{-90}{$\longrightarrow$}}
					\put(42,-13){analytic continuation}
					\put(42,-17){when $u$ crosses $1$}
				}
				\put(0,45){
					\put(0,0){\thicklines\line(1,0){10}}
					\put(30,0){\thicklines\qbezier(-20,0)(-12.5,0)(-5,-7.5)}
					\put(30,0){\thicklines\qbezier(-5,-7.5)(2.5,-15)(10,-15)}
					\put(40.5,-15){\thicklines\vector(1,0){1}}
					\put(30,0){\thicklines\qbezier(10,-15)(17.5,-15)(25,-7.5)}
					\put(30,0){\thicklines\qbezier(25,-7.5)(32.5,0)(40,0)}
					\put(70,0){\thicklines\line(1,0){5}}
					\put(72,1){$w$}
					\put(0,-10){$u>1$}
					\put(40,-10){\circle*{2}}
					\put(33,-6){$w_{*}(u,v)$}
					\put(39.5,-25){\rotatebox[origin=c]{90}{$=$}}
				}
				\put(0,10){
					\put(0,0){\thicklines\line(1,0){75}}
					\put(40,-10){\thicklines\circle{10}}
					\put(39.5,-5){\thicklines\vector(-1,0){1}}
					\put(60,0){\thicklines\vector(1,0){1}}
					\put(72,1){$w$}
					\put(40,-10){\circle*{2}}
					\put(46,-11){$w_{*}(u,v)$}
				}
			\end{picture}
		\end{tabular}
	\end{center}
	\caption{Analytic continuation of $\chi(u,v)$ across $u=1$ by deformation of the integration path in (\ref{chi hat (u,v)}).}
	\label{fig a.c.}
\end{figure}
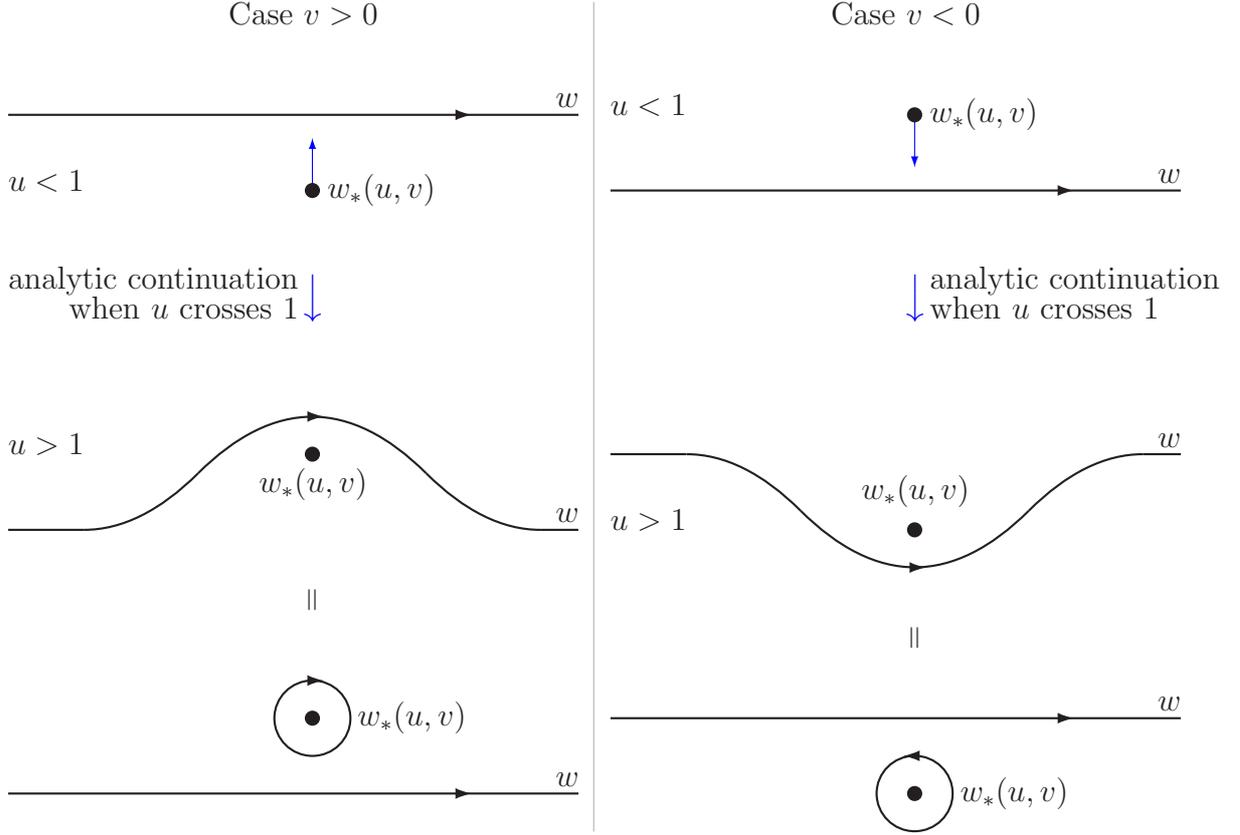

\subsection{Analytic continuation through $u=1$}
The integrand in (\ref{chi hat (u,v)}) has infinitely many poles $w_{*}(u,v)\in\mathbb{C}$, such that
\begin{equation}
	\label{poles}
	\ee^{w_{*}(u,v)^{2}/2}=u+\ii vw_{*}(u,v)\;.
\end{equation}
At $v=0$, one has for instance $w_{*}(u,0)=\pm\sqrt{2\log u+4\ii\pi n}$, $n\in\mathbb{Z}$, while at $u=0$, one finds $w_{*}(0,v)=\pm\ii\sqrt{W_{m}(v^{-2})}$, $m\in\mathbb{Z}$ with $W_{m}$ the branches of the Lambert function, such that $W_{m}(z)\,\ee^{W_{m}(z)}=z$. As explained below, however, only the purely imaginary poles $w_{*}(u,v)\in\ii\mathbb{R}$ enter the analytic continuation of $\hat{\chi}(u,v)$ for \emph{real values} of $u$ and $v$, and there are only up to three of them depending on the location in $(u,v)$-plane.

Since, as explained above, $\hat{\chi}(u,v)$ is not analytic at $u=1$ because of the pole crossing the path of integration $w\in\mathbb{R}$ at $w=0$, the analytic continuation through $u=1$ consists in adding (or subtracting depending on the direction in which the pole crosses the real axis) to $\hat{\chi}(u,v)$ the contribution of the contour integral around the pole of the integrand, i.e. $2\ii\pi$ times the residue at the pole, see figure~\ref{fig a.c.}. From (\ref{poles}), one has
\begin{equation}
	\label{w* u1}
	w_{*}(u,v)\simeq\ii\,\frac{u-1}{v}
\end{equation}
near $u=1$ for this pole, and then
\begin{equation}
	\label{chi 1+}
	\chi(u,v)=\hat{\chi}(u,v)+\delta\chi(u,v,w_{*}(u,v))
\end{equation}
with
\begin{equation}
	\label{delta chi w}
	\delta\chi(u,v,w)=\sign(v)\Big(-\ii w+\frac{uw^{2}}{2v}+\frac{\ii w^{3}}{3}\Big)
\end{equation}
for $u>1$, at least in a neighbourhood of $u=1$ and away from $v=0$. Beyond this neighbourhood, the solution $w_{*}(u,v)$ verifying (\ref{w* u1}) must be carefully followed when varying $u$ and $v$. An extra complication is however that this solution $w_{*}(u,v)$ has square root singularities on some curves in $(u,v)$-plane, at which two solutions of (\ref{poles}) coincide. As a consequence, $\chi(u,v)$ will be extended to a multivalued function, whose branches correspond to all possible exchanges of appropriate solutions of (\ref{poles}). This is however not a problem since the variables $u$ and $v$ coming from the Bethe ansatz solution do not have a physical meaning (at least to our knowledge), and $\chi$ will remain a single-valued function of the physical variables $s_{1}$ and $s_{2}$.

\subsection{Purely imaginary poles}
From the expression (\ref{delta chi w}) of the extra term $\delta\chi(u,v,w_{*}(u,v))$ in (\ref{chi 1+}) and the requirement that $\chi(u,v)\in\mathbb{R}$ for any $u,v\in\mathbb{R}$, we infer that the pole $w_{*}(u,v)$, solution of (\ref{poles}), which behaves as (\ref{w* u1}) when $u\to1$, has to be purely imaginary. Hence, we look in this section for all the solutions $w_{*}(u,v)\in\ii\mathbb{R}$ of (\ref{poles}), and then study how they are exchanged when varying $u$ and $v$.

\begin{figure}
	\begin{tabular}{ll}
		\begin{tabular}{c}\includegraphics[width=72mm]{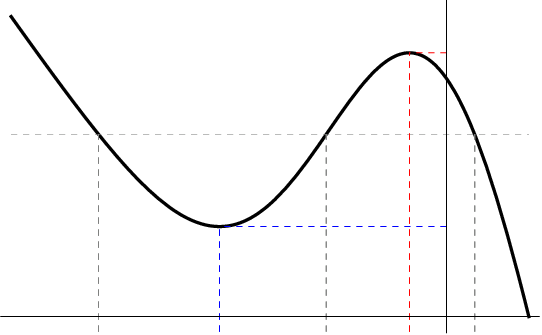}\end{tabular}
		\begin{picture}(0,0)
			\put(-60,20){$\ee^{-\omega^{2}/2}+v\omega$}
			\put(-55,13){$v<0$}
			\put(-15.5,16){\color{red}$u_{\rm max}$}
			\put(-15.5,5.3){$u$}
			\put(-15.5,-7){\color{blue}$u_{\rm min}$}
			\put(-75,-16){$\omega$}
			\put(-65.5,-23){$\omega_{\infty}$}
			\put(-50,-23){\color{blue}$\omega_{\rm min}$}
			\put(-33.5,-23){$\omega_{-}$}
			\put(-25,-23){\color{red}$\omega_{\rm max}$}
			\put(-14,-23){$\omega_{+}$}
		\end{picture}
		&
		\begin{tabular}{c}\includegraphics[width=72mm]{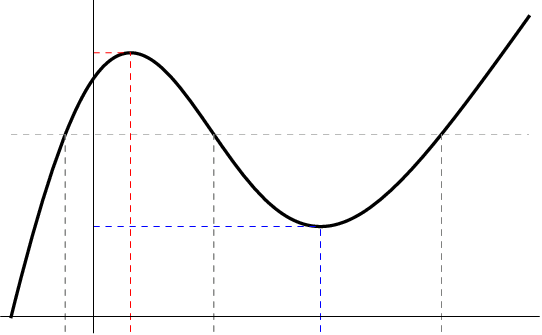}\end{tabular}
		\begin{picture}(0,0)
			\put(-45,20){$\ee^{-\omega^{2}/2}+v\omega$}
			\put(-40,13){$v>0$}
			\put(-71,16){\color{red}$u_{\rm max}$}
			\put(-66,5.3){$u$}
			\put(-70.5,-7){\color{blue}$u_{\rm min}$}
			\put(-7,-16){$\omega$}
			\put(-68,-23){$\omega_{-}$}
			\put(-60,-23){\color{red}$\omega_{\rm max}$}
			\put(-48.5,-23){$\omega_{+}$}
			\put(-37,-23){\color{blue}$\omega_{\rm min}$}
			\put(-18,-23){$\omega_{\infty}$}
		\end{picture}
	\end{tabular}
	\caption{Graph of $\ee^{-\omega^{2}/2}+v\omega$, plotted as a function of $\omega$, for $-\ee^{-1/2}<v<0$ (left) and $0<v<\ee^{-1/2}$ (right). The three distinct real solutions $\omega_{+}(u,v)$, $\omega_{-}(u,v)$ and $\omega_{\infty}(u,v)$ of (\ref{eq omega}) for $u$ such that $u_{\rm min}(v)<u<u_{\rm max}(v)$ are represented.}
	\label{fig fv+-}
\end{figure}

Writing $w_{*}(u,v)=\ii\omega$, one has
\begin{equation}
	\label{eq omega}
	\ee^{-\omega^{2}/2}+v\omega=u\;.
\end{equation}
For $v\neq0$, the left hand side, as a function of $\omega\in\mathbb{R}$, is either strictly monotonous if $|v|\geq\ee^{-1/2}$ (increasing if $v>0$, decreasing if $v<0$), or has a single local minimum and a single local maximum if $|v|<\ee^{-1/2}$, located respectively at $\omega_{\rm min}(v)=\sign(v)\,\sqrt{-W_{-1}(-v^{2})}$ and $\omega_{\rm max}(v)=\sign(v)\,\sqrt{-W_{0}(-v^{2})}$, see figure~\ref{fig fv+-}, with $W_{n}(z)$ the $n$-th branch of the Lambert function (solution of $W_{n}(z)+\log(W_{n}(z)+\ii0^{+})=\log(z+\ii0^{+})+2\ii\pi n$ where $\log$ denotes the principal branch of the logarithm with cut $\mathbb{R}^{-}$), such that $\omega_{\rm min}(v)<\omega_{\rm max}(v)<0$ if $v<0$ and $0<\omega_{\rm max}(v)<\omega_{\rm min}(v)$ if $v>0$. We define $u_{\rm min}(v)=\ee^{-\omega_{\rm min}(v)^{2}/2}+v\omega_{\rm min}(v)$ and $u_{\rm max}(v)=\ee^{-\omega_{\rm max}(v)^{2}/2}+v\omega_{\rm max}(v)$, such that $0<u_{\min}(v)<u_{\rm max}(v)$ for $|v|<\ee^{-1/2}$, see figure~\ref{fig fv+-}. Then, defining the domain in $(u,v)$-plane
\begin{equation}
	\label{T}
	T=\{(u,v),-\ee^{-1/2}\leq v\leq\ee^{-1/2},u_{\rm min}(v)\leq u\leq u_{\rm max}(v)\}\;,
\end{equation}
we observe that for any $(u,v)\notin T$, the equation (\ref{eq omega}) has a single solution, which we denote by $\omega_{*}(u,v)$. On the other hand, for any $(u,v)\in T$, (\ref{eq omega}) has three solutions, all distinct for $(u,v)$ in the interior of $T$ but with two of them coinciding on the boundary of $T$. Upon leaving $T$, the two coinciding solutions become complex valued while the third one becomes equal to $\omega_{*}$. Inside $T$, we denote the three real solutions of (\ref{eq omega}) by $\omega_{\infty}(u,v)$ and $\omega_{\pm}(u,v)$, and order them as $\omega_{-}\leq\omega_{+}\leq\omega_{\infty}$ if $v>0$ and as $\omega_{\infty}\leq\omega_{-}\leq\omega_{+}$ if $v<0$, see figure~\ref{fig fv+-}. These solutions are such that $\omega_{\pm}(u,v)$ are analytic across $v=0$ while $\omega_{\infty}$ diverges as $\omega_{\infty}(u,v)\simeq u/v$ when $v\to0$.

\begin{figure}
	\begin{center}
		\begin{tabular}{l}\includegraphics[width=75mm]{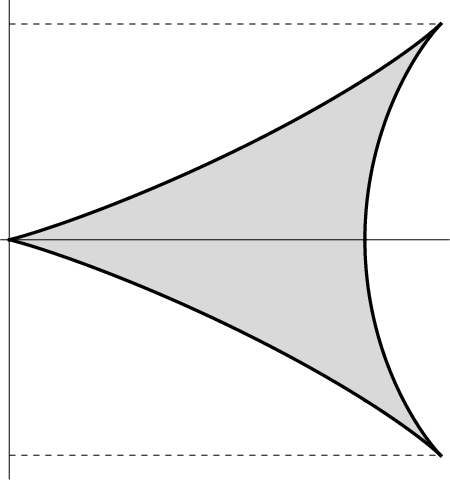}\end{tabular}
		\begin{picture}(0,0)
			\put(-5.5,3){$u$}
			\put(-75.5,40){$v$}
			\put(-86.3,36){$\ee^{-1/2}$}
			\put(-89.5,-35.5){$-\ee^{-1/2}$}
			\put(-79.5,-2){$0$}
			\put(-16.5,-2){$1$}
			\put(-65,-25){unique solution $\omega_{*}$}
			\put(-46.5,3){$3$ distinct}
			\put(-46,-1.5){solutions}
			\put(-22.5,-10){\rotatebox[origin=c]{-79}{$\omega_{+}=\omega_{-}$}}
			\put(-45,14){\rotatebox[origin=c]{26}{$\omega_{\infty}=\omega_{+}$}}
			\put(-45,-11.5){\rotatebox[origin=c]{-26}{$\omega_{\infty}=\omega_{-}$}}
			{\color{blue}
				\put(-60,-7){\vector(1,1){5}}
				\put(-62,-10){$\omega_{*}$}
				\put(-56,-1.5){$\omega_{+}$}
				\put(-60,11){\vector(1,-1){5}}
				\put(-62,12){$\omega_{*}$}
				\put(-56,3.5){$\omega_{-}$}
				\put(-13,20){\vector(-1,0){5}}
				\put(-12,18.5){$\omega_{*}$}
				\put(-22.5,18.5){$\omega_{\infty}$}
			}
		\end{picture}
	\end{center}
	\caption{Triangular domain $T$ defined in (\ref{T}), and whose boundary is given in (\ref{T edge}). The equation (\ref{eq omega}) has three distinct real solutions $\omega_{+}(u,v)$, $\omega_{-}(u,v)$, $\omega_{\infty}(u,v)$ in the interior of the triangle, and a single real solution $\omega_{*}(u,v)$ outside.}
	\label{fig triangle}
\end{figure}

Another perspective on equation (\ref{eq omega}) is that branch points $(u_{\rm b},v_{\rm b})$ where several solutions merge must verify both (\ref{eq omega}) and $\partial_{\omega}(\ee^{-\omega^{2}/2}+v\omega-u)=0$. This leads to the curve
\begin{equation}
	\label{T edge}
	u_{\rm b}(\omega)=(1+\omega^{2})\,\ee^{-\omega^{2}/2}
	\qquad\text{and}\qquad
	v_{\rm b}(\omega)=\omega\,\ee^{-\omega^{2}/2}
\end{equation}
parametrized by $\omega\in\mathbb{R}$, which has the shape of a triangle, with cusps at vertices corresponding to $\omega=\infty$ and $\omega=\pm1$, and with curved edges corresponding to the boundary of the domain $T$ defined in (\ref{T}), see figure~\ref{fig triangle}.

Writing $w_{*}(u,v)=\ii\omega_{c}(u,v)$ with a symbol $c\in\{+,-,\infty,*\}$ that must be equal to $c=*$ outside the triangle $T$ and belong to $c\in\{+,-,\infty\}$ inside $T$, the shift $\delta\chi(u,v,w_{*})$ defined in (\ref{delta chi w}) is equal to
\begin{equation}
	\label{delta chi omega}
	\delta\chi_{c}(u,v)=\sign(v)\Big(\omega_{c}(u,v)-\frac{u\,\omega_{c}(u,v)^{2}}{2v}+\frac{\omega_{c}(u,v)^{3}}{3}\Big)\;.
\end{equation}
Then, one finds that analytic continuation of the integral $\hat{\chi}(u,v)$ in (\ref{chi hat (u,v)}) from $u<1$ to $u>1$ amounts to adding the quantity $\delta\chi_{c}(u,v)$ with $c=*$ outside the triangle, and inside the triangle, $c=+$ if $v<0$ and $c=-$ if $v>0$. Further extensions of $\chi(u,v)$ across the sides of the triangle are treated in the next section.

\subsection{Full analytic continuation}
From the discussion in the previous section, the function defined as $\chi(u,v)=\hat{\chi}(u,v)$ for $u<1$ and as (\ref{chi 1+}) with $w_{*}(u,v)=\ii\omega_{*}(u,v)$ for $u>1$ is analytic outside the triangle for $(u,v)$. The situation is more complicated inside the triangle since $w_{*}(u,v)$ must be chosen among $\{\ii\omega_{+}(u,v),\ii\omega_{-}(u,v),\ii\omega_{\infty}(u,v)\}$ by continuity with $\ii\omega_{*}(u,v)$ upon entering the triangle, giving a different choice depending on the side from which the triangle is entered, respectively $\ii\omega_{+}(u,v)$, $\ii\omega_{-}(u,v)$ and $\ii\omega_{\infty}(u,v)$ when entering from below, from above, and from the right side, see figure~\ref{fig triangle}. This means that $\chi(u,v)$ can not be uniquely extended analytically within the triangle, and one must consider instead a multivalued function, whose various branches are reached by following $\hat{\chi}(u,v)$ along allowed paths in $(u,v)$-plane.

\begin{figure}
	\hspace*{-6mm}
	\begin{tabular}{ll}
		\begin{tabular}{l}
			\begin{tabular}{c}\includegraphics[width=35mm]{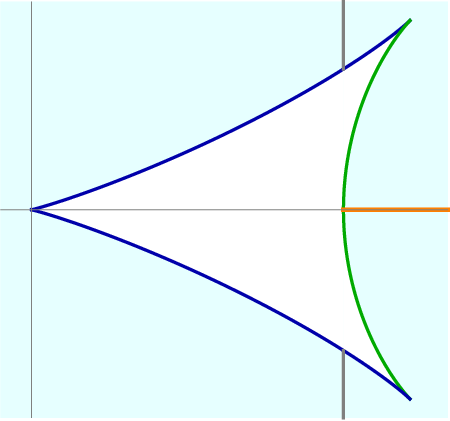}\end{tabular}
			\begin{picture}(0,0)
				\put(-6.5,3){$u$}
				\put(-35,15.5){$v$}
				\put(-25,11){$\overline{T}$}
			\end{picture}\\
			\begin{tabular}{c}\includegraphics[width=35mm]{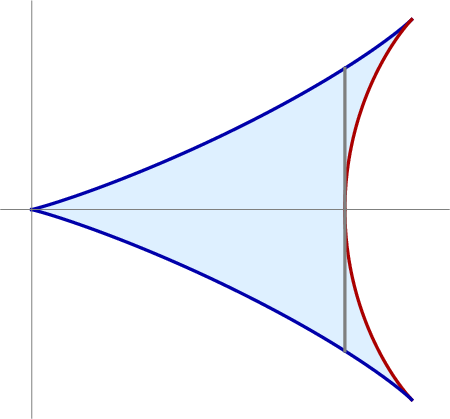}\end{tabular}
			\begin{picture}(0,0)
				\put(-6.5,3){$u$}
				\put(-35,15.5){$v$}
				\put(-20,4){$T_{1}$}
			\end{picture}\\
			\begin{tabular}{c}\includegraphics[width=35mm]{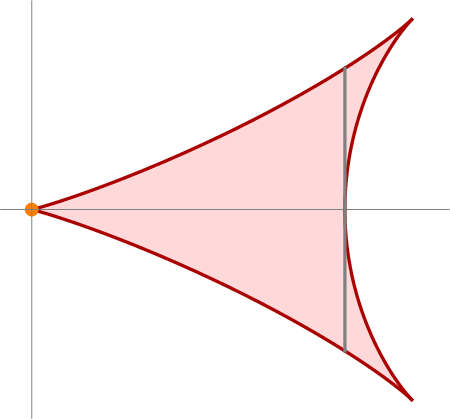}\end{tabular}
			\begin{picture}(0,0)
				\put(-6.5,3){$u$}
				\put(-35,15.5){$v$}
				\put(-20,4){$T_{2}$}
			\end{picture}\\
			\begin{tabular}{c}\includegraphics[width=35mm]{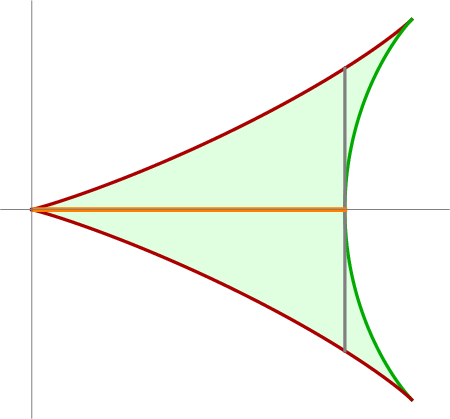}\end{tabular}
			\begin{picture}(0,0)
				\put(-6.5,3){$u$}
				\put(-35,15.5){$v$}
				\put(-20,4){$T_{3}$}
			\end{picture}
		\end{tabular}
		&
		\begin{tabular}{c}\includegraphics[width=107mm]{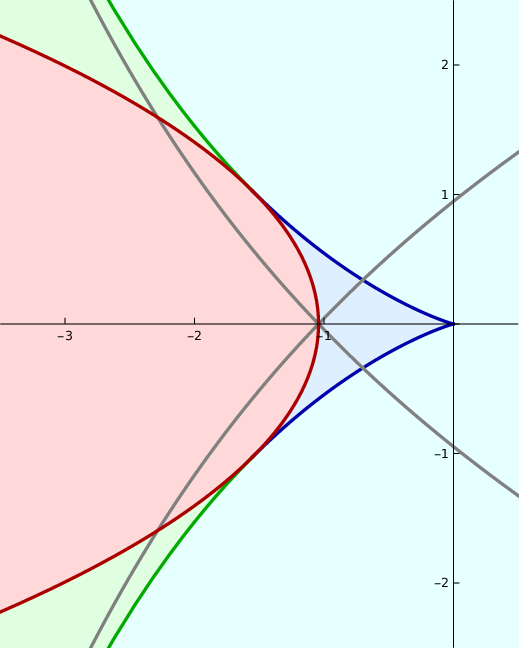}\end{tabular}
		\begin{picture}(0,0)
			\put(-7,3.5){$s_{1}$}
			\put(-16,66.5){$s_{2}$}
			\put(-12,-9){$\overline{T}$}
			\put(-40,3){$T_{1}$}
			\put(-90,-20){$T_{2}$}
			\put(-105,63){$T_{3}^{v>0}$}
			\put(-105,-61){$T_{3}^{v<0}$}
		\end{picture}
	\end{tabular}
	\caption{Domains involved in the analytic extension of $\chi$. On the left, the complement $\overline{T}$ of the triangular shape $T$ defined in (\ref{T}) and the three copies $T_{i}$, $i=1,2,3$ of $T$, for which $\chi$ is respectively given in (\ref{chi a.c. out}), (\ref{chi a.c. in 1}), (\ref{chi a.c. in 2}), (\ref{chi a.c. in 3}), are represented in $(u,v)$-plane. The function $\chi$ goes to infinity when approaching the orange portions (half-line $u\geq1$, $v=0$ for $\overline{T}$ ; point $u=v=0$ for $T_{2}$ ; segment $0\leq u\leq1$, $v=0$ for $T_{3}$). The colour scheme for the edges of the triangle indicates identification of the corresponding edges under analytic continuation. On the right, the same domains are represented in $(s_{1},s_{2})$-plane, with the same colours.}
	\label{fig domains T}
\end{figure}

Two types of obstructions forbid some paths in $(u,v)$-plane. First, $\chi(u,v)$ has a divergent contribution in some regions of the axis $v=0$ : the integral (\ref{chi hat (u,v)}) if $u>1$, terms containing $\omega_{\infty}(u,v)$ (which is divergent at $v=0$), or the term proportional to $v^{-1}$ in (\ref{delta chi omega}) (although that divergence ends up cancelling with another one in some cases). Obstructions may also happen when, starting from within the triangle, two real solutions of (\ref{eq omega}) among $\omega_{+}$, $\omega_{-}$, $\omega_{\infty}$ coincide at the edge of the triangle with a square root singularity. If the expression for $\chi(u,v)$ contains either of the solutions coinciding at the edge, one can not cross that edge, but must instead go back inside the triangle after exchanging the two solutions, which leads to an expression for $\chi$ locally analytic in e.g. variables $s_{1}$ and $s_{2}$.

Working out the analytic extensions induced by crossing the line $u=1$ and the edges of the triangle, we finally obtain the single determination
\begin{equation}
	\label{chi a.c. out}
	\fl
	\chi(u,v)\underset{(u,v)\notin T}{=}
	\left\{\!\!
	\begin{array}{ll}
		\hat{\chi}(u,v) & u<1\\
		\hat{\chi}(u,v)+\delta\chi_{*}(u,v) & u>1
	\end{array}
	\right.
\end{equation}
for $(u,v)$ outside the triangle of figure~\ref{fig triangle}, with $\hat{\chi}(u,v)$ defined in (\ref{chi hat (u,v)}), see also (\ref{chi hat (u)}) for the limit $v\to0$, and $\delta\chi_{*}(u,v)$ in (\ref{delta chi omega}), with $\omega_{*}(u,v)$ the unique real solution of (\ref{eq omega}). For $(u,v)$ inside the triangle, one has three determinations for $\chi(u,v)$ given respectively by
\begin{eqnarray}
	\label{chi a.c. in 1}
	&&\fl
	\chi_{1}(u,v)=
	\left\{\!\!
	\begin{array}{ll}
		\hat{\chi}(u,v) & \text{if}\;u<1\\
		\hat{\chi}(u,v)+\delta\chi_{-}(u,v) & \text{if}\;u>1\;\text{and}\;v>0\\
		\hat{\chi}(u,v)+\delta\chi_{+}(u,v) & \text{if}\;u>1\;\text{and}\;v<0
	\end{array}
	\right.\;,\\
	\label{chi a.c. in 2}
	&&\fl
	\chi_{2}(u,v)=
	\left\{\!\!
	\begin{array}{ll}
		\hat{\chi}(u,v)+\sign(v)(\delta\chi_{+}(u,v)-\delta\chi_{-}(u,v)) & \text{if}\;u<1\\
		\hat{\chi}(u,v)+\delta\chi_{+}(u,v) & \text{if}\;u>1\;\text{and}\;v>0\\
		\hat{\chi}(u,v)+\delta\chi_{-}(u,v) & \text{if}\;u>1\;\text{and}\;v<0\\
	\end{array}
	\right.\;,\\
	\label{chi a.c. in 3}
		&&\fl
	\chi_{3}(u,v)=
	\left\{\!\!
	\begin{array}{ll}
		\hat{\chi}(u,v)+\delta\chi_{\infty}(u,v)-\delta\chi_{-}(u,v) & \text{if}\;u<1\;\text{and}\;v>0\\
		\hat{\chi}(u,v)+\delta\chi_{\infty}(u,v)-\delta\chi_{+}(u,v) & \text{if}\;u<1\;\text{and}\;v<0\\
		\hat{\chi}(u,v)+\delta\chi_{\infty}(u,v) & \text{if}\;u>1
	\end{array}
	\right.\;,
\end{eqnarray}
with $\delta\chi$ defined in (\ref{delta chi omega}) from the real solutions $\omega_{\pm}$ and $\omega_{\infty}$ of (\ref{eq omega}) specified as in figure~\ref{fig fv+-}. The functions $\chi_{1}$, $\chi_{2}$ and $\chi_{3}$ are analytic in the interior of the triangle, except at $v=0$ for $\chi_{3}$, which is divergent there.

We emphasize that, while the various branches of the function $\chi$ are not analytic functions of $u$ and $v$ at the boundaries of the triangle, because of square root branch points, these branches induce a single valued, analytic function of the variables $s_{1}$ and $s_{2}$ defined in (\ref{s1[chi]}) and (\ref{s2[chi]}), see figure~\ref{fig domains T}.

\vspace{10mm}

%\bibliographystyle{unsrt}
%\bibliography{/users/prolhac/bib/references.bib}

\end{document}